# A 64-Spin All-to-All CMOS Ising Machine with Landscape Perturbation Achieving 2.28 nJ/Edge-Bit Energy-to-Solution

Ahmet Yusuf Salim*[1], Jianan Wu*[1], Soner Seckiner[1], Eslam Elmitwalli[1], Selçuk Köse[1], and Zeljko Ignjatovic[1]

1. Department of Electrical and Computer Engineering, University of Rochester, 160 Trustee Rd, Rochester, NY 14627, USA

jwu144@ur.rochester.edu

*These authors contributed equally to this work.

*Abstract*—A 64-spin all-to-all current-mode coupling Ising machine is implemented in 65 nm CMOS. The design supports 31 coefficient levels in 0.943 mm$^2$ and achieves Energy-to-Solution (ETS) of 2.28 nJ/edge-bit. Continuous programming refresh not only mitigates leakage but also provides a mechanism for deterministic energy landscape perturbation, which consistently improves solution quality with higher success rate compared to operation without landscape perturbation.

I. INTRODUCTION

Combinatorial optimization problems (COPs) such as Max-Cut, graph coloring and graph partitioning are fundamental in scientific and engineering applications. These problems are NP-hard, and the computational cost of finding optimal solutions grows rapidly with problem size, making conventional von Neumann computing inefficient. A unified way to express such problems is Quadratic Unconstrained Binary Optimization (QUBO), which maps directly to the Ising Hamiltonian, where binary spin variables interact via programmable couplings. CMOS-based Ising machines (IMs) have attracted considerable interest due to their potential for room-temperature, low-cost integration [1-6]. In particular, continuous-time Ims [2-6], have been widely studied because their intrinsic parallelism allows all spins to evolve simultaneously toward energy minima, enabling fast convergence. Prior demonstrations include ring oscillator-based designs that realize all-to-all connectivity through phase coupling [2,3], as well as planar architectures that reduce circuit complexity but lack full connectivity [4-6]. These implementations have successfully demonstrated the feasibility of CMOS IMs, yet challenges remain in achieving power efficiency, scalability and solution quality. In particular, oscillator-based designs suffer from high static power, large area overhead and strong sensitivity to mismatch and phase noise [2,3], while planar topologies require embedding into restricted connectivity graphs [4-6], introducing overhead and reducing mapping accuracy. Moreover, most CMOS IMs lack explicit perturbation mechanisms, limiting their ability to escape local minima and achieve higher-quality solutions [2,3,5,6]. These limitations motivate the need for an IM that offers power and area efficiency and higher solution quality, and operates without embedding.

In this work, we present a continuous-time all-to-all current-mode IM that addresses these challenges. A qualitative feature-level comparison with prior IMs is provided in Table I and the key features of the proposed IM is illustrated in Fig. 1. Its native all-to-all connectivity removes the need for embedding, allowing accurate and low-overhead mapping of QUBO problems. Realizing this all-to-all coupling requires two main components: coupling polarity sgn($J_{ij}$) and coupling strength $|J_{ij}|$. In the circuit, polarity is implemented through current sourcing or sinking, while the coupling strength is set through programmable biasing of each unit achieved with current-steering digital-to-analog converters (DACs). With a 4-bit resolution per unit, plus a sign bit, the coupling can be programmed to 31 distinct levels. Additionally, spins are represented as capacitor voltages and read out by inherently linear 1-bit inverter-based ADCs, ensuring compact and power-efficient implementation.

TABLE I. Comparison with Prior Works

| | This Work | Ring Oscillator Ising [3] | Planar Ising [5, 6] |
|---|---|---|---|
| Computing Type | Continuous-time (one-shot), fully parallel 😊 | | |
| Connectivity | All-to-all 😊 | | Lattice (requires embedding) 😞 |
| Robustness | Robust to leakage 😊 | Sensitive to mismatch, phase noise 😞 | Sensitive to mismatch 😞 |
| Annealing | Landscape perturbation 😊 | None (noise only) 😞 | None (noise only) 😞 |
| Time-to-Solution | Short 😊 | Long 😞 | N/A |
| Scalability | Area-efficient, scalable 😊 | Poor (oscillator area) 😞 | Poor (embedding overhead) 😞 |

The proposed landscape perturbation scheme arises from the way the coupling bias is generated. Each of the 64 nodes is coupled to all other 63 nodes, with the coupling units (CUs) arranged in a matrix. Programming DACs apply the bias voltage by sequentially activating switches column by column, after which the voltages are stored on the gate capacitors of CUs. In the proposed design, charge leakage at the gate of CUs gradually distorts the programmed coefficients, degrading solution quality. To address this issue, a continuous programming scheme is employed, in which the coupling matrix is periodically refreshed in a column-wise manner. This mechanism not only mitigates the impact of leakage by maintaining the effective coupling values but also enables a unique landscape perturbation process.

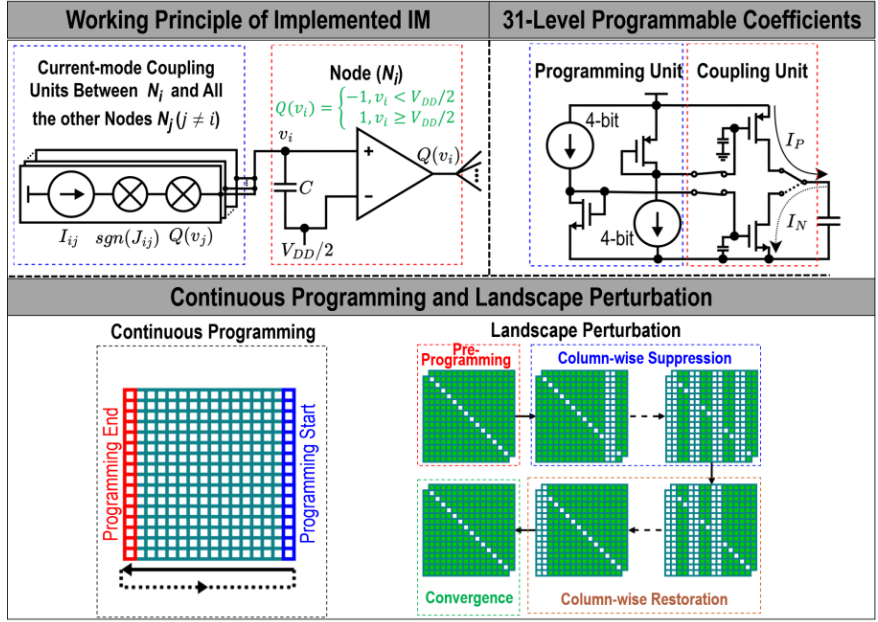

Fig. 1. Key features of this work. Working principle of implemented IM. (top left), 4-bit DAC with 1-bit sign control for programmable coupling units (top right), and continuous programming and landscape perturbation (bottom).

## II. Proposed 64-Spin All-to-All IM

### A. Ising Model and QUBO Problems

The Ising model provides an energy-based representation for optimization problems. The general form can be written as

$$H_{\text{Ising}} = -\sum_{i<j} J_{ij}\sigma_i\sigma_j - \sum_i h_i\sigma_i \tag{1}$$

where $\sigma_i \in \{-1, +1\}$ is the spin state, $J_{ij}$ encodes the interaction strength, and $h_i$ denotes the bias field.

Many QUBO problems can be mapped directly to the bias-free Ising form. For example, Max-Cut can be written in QUBO form by assigning a binary variable $x_i \in \{0,1\}$ to each vertex and representing an edge $(i,j)$ with weight $W_{ij}$ being cut through the quadratic term $W_{ij}(x_i + x_j - 2x_i x_j)$. Substituting $\sigma_i = 2x_i - 1$, the Max-Cut objective becomes $\max\left(const - \frac{1}{2}\sum_{i<j} W_{ij}\sigma_i\sigma_j\right)$, which is equivalent to minimizing the bias-free Ising Hamiltonian

$$H = -\sum_{i<j} J_{ij}\sigma_i\sigma_j, \qquad J_{ij} = -W_{ij} \tag{2}$$

*B. Dynamics of proposed IM*

To explain the dynamics of the proposed IM, we derive the governing differential equation of a node and the corresponding energy function from Fig. 1 (top left). The input current to each node is determined by $\text{sgn}(J_{ij})Q(v_j)$, scaled by the coupling current $I_{ij}$, where $Q(v_j)$ denotes the spin state of node $j$. This leads to the node-update equation

$$\frac{dv_i}{dt} = \frac{1}{C}\sum_{j \neq i} \text{sgn}(J_{ij}) I_{ij} Q(v_j) \tag{3}$$

Since $I_{ij} = a|J_{ij}|$ with $a > 0$, the dynamics can be rewritten as

$$\frac{dv_i}{dt} = \frac{a}{C}\sum_{j \neq i} J_{ij} Q(v_j) \tag{4}$$

The system evolves according to

$$H(v) = -\sum_{i<j} J_{ij} Q(v_i) Q(v_j), \quad Q(v_i) = \begin{cases} -1 & v_i < V_{DD}/2 \\ +1 & v_i \geq V_{DD}/2 \end{cases} \tag{5}$$

which is equivalent to the bias-free Ising Hamiltonian, where $V_{DD}$ denotes power supply voltage. Differentiating $H$ with respect to time is

$$\begin{aligned}
\frac{dH(v)}{dt} &= \sum_i \frac{\partial H(v)}{\partial Q(v_i)} \frac{dQ(v_i)}{dt} \\
&= \sum_i \left[ -\left(\sum_{j \neq i} J_{ij} Q(v_j)\right) \frac{dQ(v_i)}{dt} \right] \\
&= -\frac{C}{a}\left[\frac{dv_i}{dt} \frac{dQ(v_i)}{dt}\right]
\end{aligned} \tag{6}$$

Since $\frac{dv_i}{dt}$ and $\frac{dQ(v_i)}{dt}$ always change in the same direction, the energy is non-increasing over time, and the system naturally converges toward a low-energy solution.

*C. Overall Architecture and Workflow*

The overall architecture and workflow are illustrated in Fig. 2. The operation of the IM begins with writing interaction sign and coupling strength values (5 bits per CU) into the register-based memory via an 8-bit bus. Once the memory is loaded, spins are initialized by a 64-bit linear feedback shift register (LFSR), where each output drives one spin, forming a 64-bit configuration. An external digital controller generates rising-edge pulses (CLK$_{INIT}$) that shift the LFSR by one bit per solve, producing distinct configurations with negligible power overhead. Next, the CUs are programmed by transferring data from the memory to row-level 5-bit DACs, which generate bias voltages for current sources in the selected CU column. A 64-bit shift register located between the memory and coupling array, operating at 80 MHz, activates the appropriate memory cells and CU column, repeating 64 times until the array is fully programmed. After programming, the system anneals for 3 µs, evolving toward low-energy states corresponding to the

This work was supported in part by NSF under Award No. 2233378 and by DARPA under contract No. FA8650-23-C-7312.

problem solution, while the column DAC continuously refreshes the bias voltages of CUs. At the end of annealing, spin states are latched into node registers and read out sequentially off-chip, completing one run.

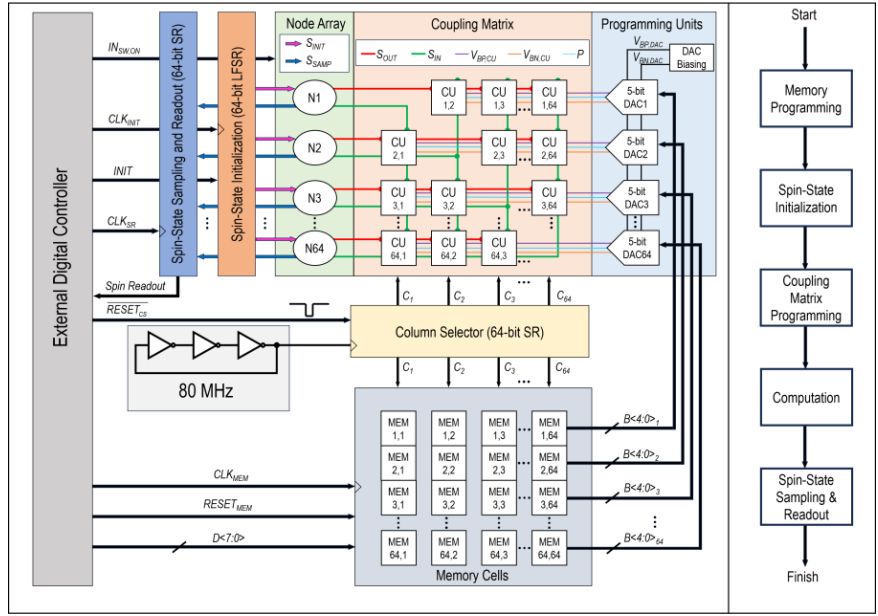

Fig. 2. Proposed IM architecture with key design block (left) and workflow (right).

### D. Schematic and Layout of the Key Blocks

Fig. 3 illustrates the building blocks of the proposed hardware and the layout of the active area. The CU on top left incorporates a 6 transistor SRAM cell for storing the sign bit and logic for generating polarity-dependent currents that drive the node voltages. The node circuit on top right contains spin initialization logic and capacitor-voltage based spin information storage, which is digitized through a series of inverters. Additionally, nodes enable the current coupling by integrating the charge/discharge contribution from all CUs connected to the node. The current steering DAC on the bottom left programs the coupling strength with 4-bit resolution and sign control, and transfers bias voltages to the gates of current sources and sinks within the CUs. Finally, the layout on bottom right integrates these elements into a compact design, consisting of a 64×64 coupling matrix flanked by node and DAC arrays, occupying an area of 560 μm × 314 μm.

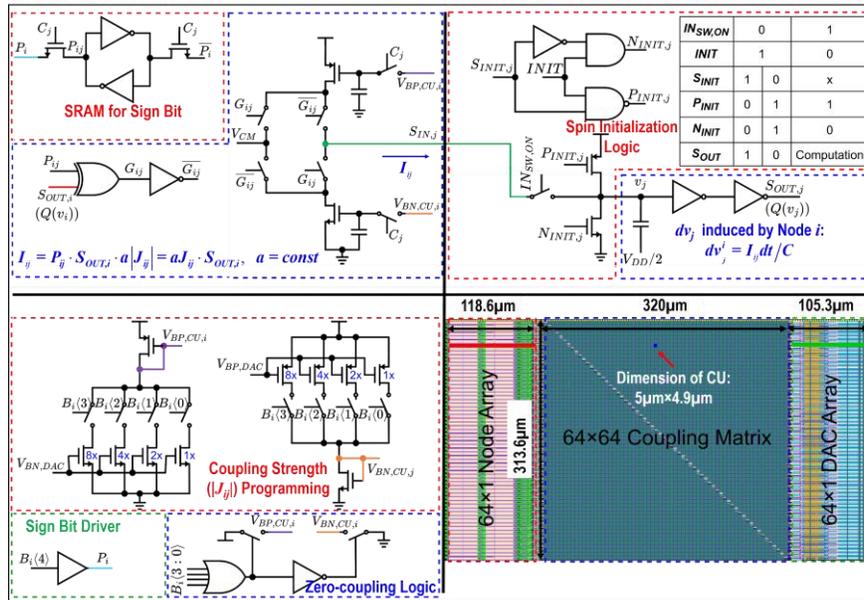

Fig. 3. Schematics of the coupling unit (top left), node (top right), DAC (bottom left), and their corresponding layout (bottom right).

## III. COUPLING PROGRAMMING AND LANDSCAPE PERTURBATION

Continuous programming with landscape perturbation is illustrated in Fig. 1 (bottom). Row-level DACs sequentially refresh the coupling columns, with the refresh cycle periodically restarted by a reset signal. In nominal mode, constant biases ($V_{BP,DAC}$ and $V_{BN,DAC}$) keep both DAC branches active, so the selected column is refreshed to its programmed values. For perturbation, the DAC biases are periodically gated off; during the disable window, the currently selected column is forced to zero. Accordingly, CUs can be categorized as programmed, disabled (zeroed), and restored (reprogrammed later). This periodic column-wise suppression temporarily reduces coupling density and helps the system escape shallow local minima; subsequent refresh restores the original Hamiltonian for final convergence. The scheme also compensates leakage and provides deterministic landscape modulation, improving the probability of reaching the global minimum.

Fig. 4 (left) shows simulated energy trajectories for a 64-node randomly generated QUBO problem under landscape perturbation (solid) and without perturbation using gradient descent only (dashed), with two initial spin configurations (red and green). Without perturbation, the Hamiltonian decreases monotonically and quickly gets trapped in local minima. With landscape perturbation, the energy fluctuates during coupling suppression and restoration periods, allowing both trajectories to escape shallow minima and converge to the global optimum. These simulation results provide an intuitive illustration of how landscape perturbation reshapes the energy landscape during the annealing. Following prior work [3], we consider a run successful if its Hamiltonian achieves at least 99% of the best-known energy obtained via Tabu-based software solver [7]. Fig. 4 (right) compares measured and simulated Hamiltonian histograms relative to the 99% best-known energy threshold. The measured results show that inherent perturbation (inherent noise and other circuit non-idealities) and the simulated baseline without any perturbations (i.e., gradient descent alone) achieve similar success rates (SRs), while the proposed landscape perturbation improves the SR by more than 1.7× compared to both cases.

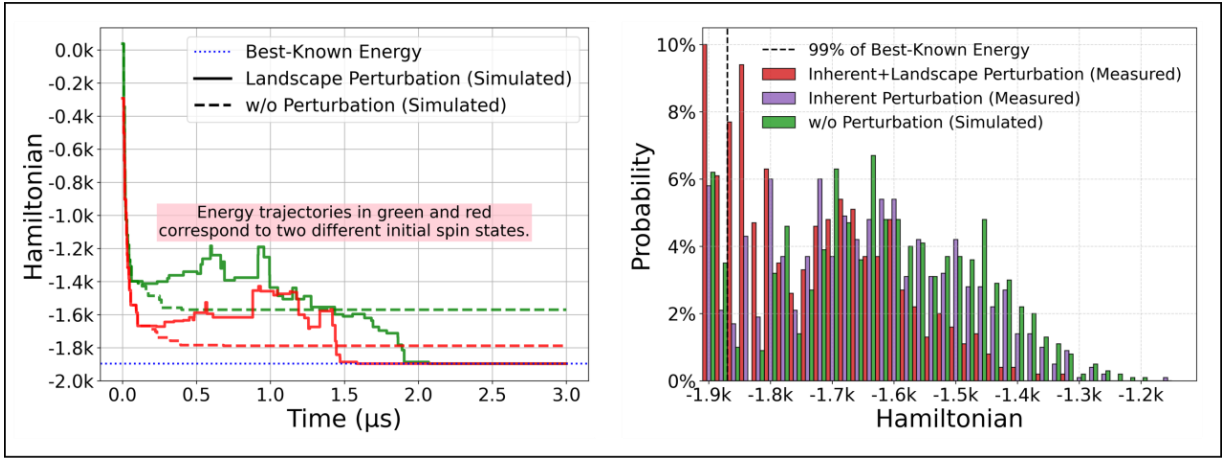

Fig. 4. Simulated Hamiltonian trajectories for a randomly generated 64-node QUBO problem (left), and measured vs. simulated Hamiltonian histograms (right).

## IV. MEASUREMENT RESULTS

Fig. 5 (top) summarizes measured SRs under landscape perturbation across problem sizes from 16 to 64 nodes and problem densities from 10% to 90% with each coupling coefficient chosen at random from –15 to +15. Each QUBO problem is solved 1000 times to evaluate its SR. For each size–density pair, the mean across 20 random problems is plotted. The figure suggest that SR tends to decrease as problem size increases, which is expected and consistent with larger instances creating more complex energy landscapes and a higher likelihood of being trapped in local minima. SR also shows an upward trend with density in our measurements, a trend that has likewise been observed in ring-oscillator-based IMs [2,3]. One possible explanation is that in low-density problems, the small number of couplings makes each edge more influential, resulting in a more discretized Hamiltonian landscape with deeper local minima. Under such conditions, circuit non-idealities such as mismatch and leakage also have a larger relative impact, further reducing solution quality. Fig. 5 (bottom) shows the cumulative distribution of Time-to-Solution (TTS) for the same 100 64-node problems in Fig. 5 (top). TTS is defined as the time required to reach a target success probability (e.g., 99%) and is computed as

$$TTS = \tau \frac{\ln(0.01)}{\ln(1 - p_{suc})} \qquad (7)$$

Where $\tau$ is the per-run anneal time and $p_{suc}$ is the single-run success rate [8]. The curve shows the number of problems solved versus TTS, with the mean (1.56 ms, red) and median (0.72 ms, green) indicated.

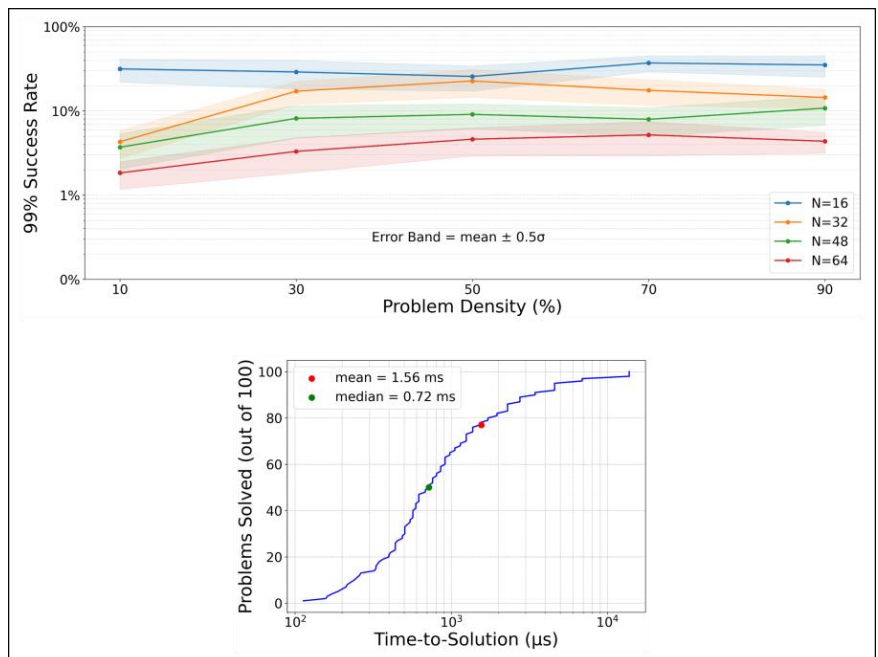

Fig. 5. 99% success rates for 400 randomly generated QUBO problems of various sizes and densities with landscape perturbation (top) and Time-to-Solution (TTS) cumulative distribution for 100 64-node randomly generated problems (bottom).

Table II presents a table of state-of-the-art IMs reported in various venues. The proposed hardware outperforms comparable works in the key metrics TTS and ETS by a wide margin. Fig. 6 shows the chip micrograph and summarizes the hardware metrics, while Fig. 7 presents the testbench.

TABLE II. COMPARISON OF THE PROPOSED IM WITH STATE-OF-THE-ART WORKS

|  | This Work | Nat.Electron.'23 [2] | Nat.Electron.'25 [3] | JSSC'24 [4] | ISSCC'23 [5] | ISSCC'24 [6] |
|---|---|---|---|---|---|---|
| Technology | 65nm CMOS | | | | | |
| Spin State Representation | Capacitor Voltage | Ring-Oscillator Phase | Ring-Oscillator Phase | Capacitor Voltage | Latch Phase | Latch Phase |
| Spin Interaction | Current | Transmission Gate | Logic Gate | Current | Switches | Switches |
| Graph Topology (Graph Directionality) | All-to-All (Directed) | All-to-All (Undirected) | All-to-All (Undirected) | King (Directed) | Lattice (Undirected) | Lattice (Undirected) |
| Number of Spins (Interaction Number) | 64 (63) | 48 (47) | 59 (58) | 1056 (8) | 1440 (4) | 2304 (4) |
| Coeff. Level | 31 | 29 | 29 | 15 | 3 | 3 |
| Core Area (mm²) | 0.943 | 1.8 | 4.8 | 3.2 | 0.446 | 0.611 |
| Normalized Spin Area[A] (μm²) | 23.5 | 164.2 | 288.7 | 13.6 | 48.9 | 41.8 |
| Power (mW) | 31.6@1.2 V[B] | 105@1.2 V | 10@1.2 V | 50.7@1.2 V | 150@1.05 V | 240@1.2 V |
| Anneal Time | 3 μs | <14 μs | <100 μs | <20.7 ns | <20 ns | <100 ns |
| TTS (ms) | 0.72 | N/A | <15.12 | N/A | N/A | N/A |
| ETS[C] (μJ) | 22.76 | N/A | <151.2 | N/A | N/A | N/A |
| Normalized ETS[D] (nJ) | 2.28 | N/A | <18.19 | N/A | N/A | N/A |

[A] Normalized Spin Area = $\dfrac{\text{Spin area}}{\log_2(\text{Coeff. level}) \cdot (\text{Graph directionality}) \cdot (\text{Interaction number})}$, Graph directionality = 2 for directed and 1 for undirected [4].

[B] Includes all on-chip components.

[C] ETS = Power · TTS

[D] Normalized ETS = $\dfrac{\text{ETS}}{\log_2(\text{Coeff. level}) \cdot (\text{Number of edges})} = \dfrac{\text{ETS}}{\log_2(\text{Coeff. level}) \cdot (\text{Number of spins}) \cdot (\text{Interaction number}) / 2}$

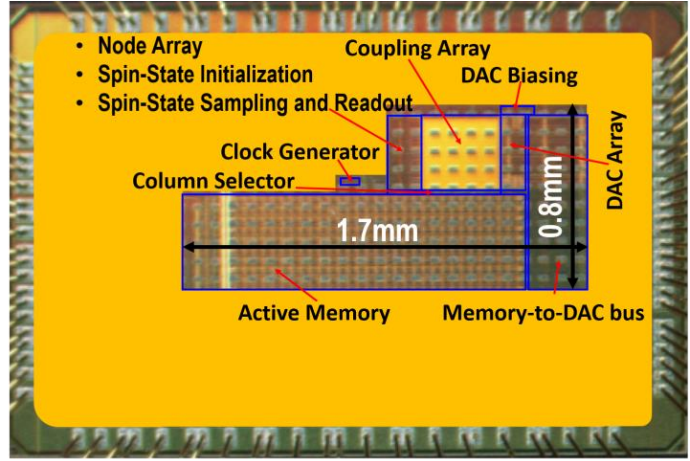

| Technology | 65 nm CMOS |
|---|---|
| Design Highlights | • 64-node All-to-all Current-Mode Coupling<br>• 31-level Coupling Coefficients<br>• Continuous Programming Enabling Leakage Tolerance<br>• Landscape Perturbation for Performance Boost<br>• Area- and Power-Efficient |
| Core Size | 0.943 mm² |
| TTS | 0.72 ms |
| ETS | 22.76 µJ |
| Normalized Spin Area | 23.5 µm² |
| Normalized ETS | 2.28 nJ |

Fig. 6. Die micrograph and summary table.

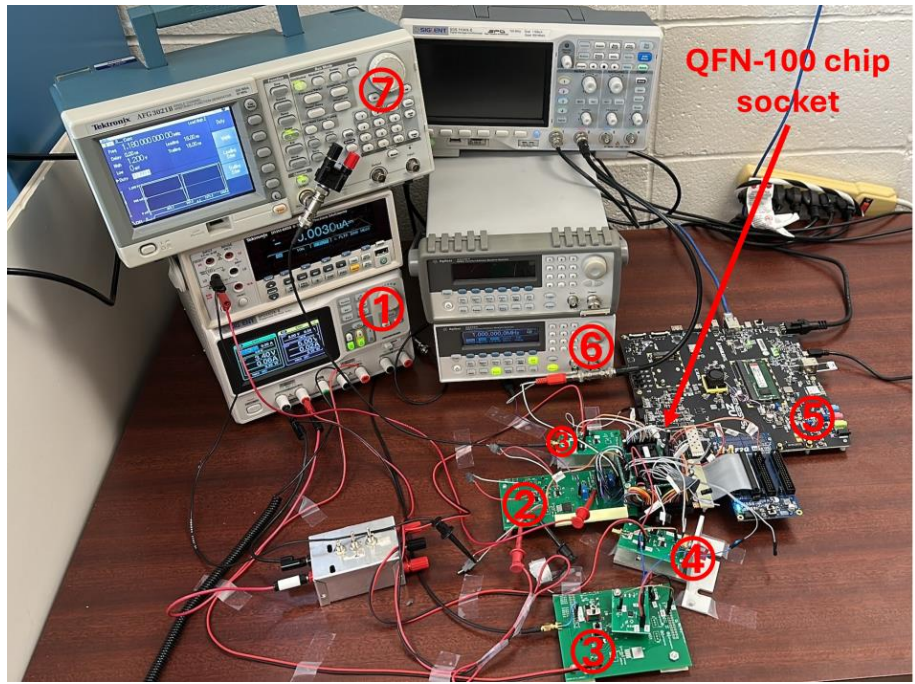

Fig. 7: Experimental measurement setup of the proposed system. (1) Main power supply providing energy to the measurement environment. (2) Power supply/interface board. (3) Voltage level shifters interfacing the chip and function generators. (4) Annealing-time fine-tuning circuitry enabling adjustment of operation period precisely. (5) External digital controller board for configuration and data transfer. (6) Function generator used to control the DAC biasing module by switching its supply between $V_{\text{DD}}$ (positive supply) and $V_{SS}$ (ground), thereby enabling or disabling the DAC. (7) Function generator used to control the coupling refresh rate.

ACKNOWLEDGMENT

This work was supported in part by NSF under Award No. 2233378 and by DARPA under contract No. FA8650-23-C-7312. We would like to thank Mathew X. Burns, Bryan Jacobs, John Berberian, Yeuan-Ming Sheu, and Jeff Stone for their valuable feedback.